\documentclass[twocolumn,showpacs,superscriptaddress,aps,prb]{revtex4-1}
\usepackage{bm}
\usepackage{amssymb,amsmath}
\usepackage[dvips]{graphicx}
\usepackage[colorlinks,linkcolor=blue,citecolor=blue,urlcolor=blue]{hyperref}

\begin{document}

\title{Rounding of a first-order quantum phase transition to a strong-coupling critical point}

\author{Fawaz Hrahsheh}
\affiliation{Department of Physics, Missouri University of Science and Technology, Rolla, Missouri 65409, USA}

\author{Jos\'e A. Hoyos}
\affiliation{Instituto de F\'{\i}sica de S\~ao Carlos, Universidade de S\~ao Paulo,
C.P. 369, S\~ao Carlos, S\~ao Paulo 13560-970, Brazil}

\author{Thomas Vojta}
\affiliation{Department of Physics, Missouri University of Science and Technology, Rolla, Missouri 65409, USA}

\begin{abstract}
We investigate the effects of quenched disorder on first-order quantum
 phase transitions on the example of the $N$-color quantum Ashkin-Teller model.
By means of a strong-disorder renormalization group, we demonstrate
that quenched disorder rounds the first-order quantum phase transition to a continuous one for both
 weak and strong coupling between the colors. In the strong coupling case, we find a distinct
type of infinite-randomness critical 
point characterized by additional internal degrees of freedom.
We investigate its critical properties in detail and find stronger thermodynamic singularities than in 
the random transverse field Ising chain. We also discuss the implications for higher spatial dimensions as well 
as unusual aspects of our renormalization-group scheme.

\end{abstract}

\date{\today}

\maketitle
 \section{Introduction}
\label{Intr}
The effects of disorder on quantum phase transitions have gained increasing attention 
recently, in particular since experiments
have discovered several of the exotic phenomena predicted by theory (see, e.g., Refs. 
\onlinecite{Vojta_review06, Vojta_review10}). Most of the existing work
has focused on continuous transitions while first-order quantum phase transitions have 
received less attention.  
 
In contrast, the influence of randomness on pure systems undergoing a \emph{classical} 
first-order transition has been comprehensively studied. 
Using a beautiful heuristic argument, Imry and Wortis \cite{ImryWortis79} reasoned that 
quenched disorder should round 
classical first-order phase transitions in sufficiently low dimension and thus produce 
new continuous phase transitions.
This analysis was extended by Hui and Berker.\cite{HuiBerker89} 
Aizenman and Wehr \cite{AizenmanWehr89} rigorously proved that first-order phase transitions
cannot exist in disordered systems in dimensions $d\le 2$. If the randomness
 breaks a continuous symmetry, the marginal dimension is $d=4$. 

The question of whether or not disorder can round a first-order {\it quantum}
phase transition (QPT) to a continuous one was asked by
Senthil and Majumdar,\cite{SenthilMajumdar96} and, more recently,
by Goswami \textit{et al}.\cite{GoswamiSchwabChakravarty08} 
Using a strong-disorder renormalization group (SDRG) technique, they found that the transitions
in the random quantum Potts and clock chains~\cite{SenthilMajumdar96} were governed by the well-known infinite-randomness critical point
(IRCP) of the random transverse-field Ising chain.\cite{Fisher92, Fisher95} The same holds for the $N$-color
 quantum Ashkin-Teller (AT) model in the weak-coupling (weak interaction between the colors) regime.\cite{GoswamiSchwabChakravarty08}
This implies that disorder can indeed round first-order quantum phase transitions.

In the strong-coupling regime of the AT model, on the other hand, the 
renormalization-group (RG) analysis of the authors of
Ref.~\onlinecite{GoswamiSchwabChakravarty08} breaks down.  
Goswami \textit{et al.} speculated that 
this implies persistence of the first-order QPT in the presence of disorder,
requiring important modifications
of the Aizenman-Wehr theorem. 
However, shortly after, Greenblatt \textit{et al.} 
\cite{GreenblattAizenmanLebowitz09,GreenblattAizenmanLebowitz12} proved
rigorously that the Aizenman-Wehr theorem also holds for quantum systems at zero 
temperature.

\begin{figure}
\includegraphics[width=1\columnwidth]{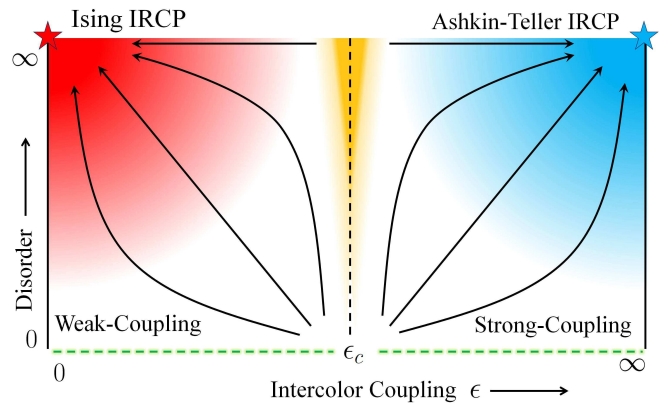}
\caption{(Color online) Schematic of the renormalization-group flow diagram 
in the disorder---coupling strength parameter space.
For $\epsilon<\epsilon_c$ (left arrows), the critical flow approaches the 
usual Ising infinite-randomness critical point of Ref.~\onlinecite{Fisher95}.
For $\epsilon>\epsilon_c$ (right arrows), we find a distinct infinite-randomness
 critical point with even stronger thermodynamic singularities.
}
\label{FlowDiagram}
\end{figure} 

In this paper, we resolve the apparent contradiction between these results.
 We show that quenched disorder rounds the first-order QPT of the AT model in the
strong-coupling regime as well as in the weak-coupling regime.
Moreover, we unveil a distinct type of infinite-randomness critical point
governing the transition in the strong-coupling regime. 
It is characterized by additional internal degrees of freedom which appear
 because a higher symmetry is dynamically generated at criticality. As a consequence, 
the critical point displays even stronger thermodynamic singularities than the transverse-field Ising IRCP.
To obtain these results, we have developed an implementation of the SDRG method
that works for both weak and strong coupling. 
In particular, this method can deal with the diverging
intercolor interactions as well as the associated additional degeneracies.
A schematic of the resulting RG flow in the critical plane is shown in Fig.~\ref{FlowDiagram}.

Our paper is organized as follows: In Sec.~\ref{QATM}, we define the model and discuss a few of its basic properties. Section~\ref{SDRG} is 
devoted to our strong-disorder renormalization group scheme. The resulting phase diagram and observables are discussed in Sec.~\ref{PDO}.
We conclude in Sec.~\ref{Conc}.

\section{Quantum Ashkin-Teller model}
\label{QATM}
The Hamiltonian of the one-dimensional $N$-color quantum AT 
model~\cite{GrestWidom81, Fradkin84, Shankar85} is given by
\begin{equation}
\label{H_label}
\begin{gathered}
 H=-\sum_{\alpha=1}^N\sum_{i=1}^L{\left ( J_i \sigma_{\alpha,i}^z \sigma_{\alpha,i+1}^z
+ h_i \sigma_{\alpha,i}^x \right )}\\
-\sum_{\alpha<\beta}^N\sum_{i=1}^L{\left (\epsilon_{J,i}J_i \sigma_{\alpha,i}^z 
\sigma_{\alpha,i+1}^z \sigma_{\beta,i}^z \sigma_{\beta,i+1}^z
+ \epsilon_{h,i}h_i \sigma_{\alpha,i}^x \sigma_{\beta,i}^x\right )}.
\end{gathered}
\end{equation}
Here, $i$ indexes the lattice sites, $\alpha$ and $\beta$ index colors, and $\sigma^x$
 and $\sigma^z$ are the usual Pauli matrices. 
The interactions $J_i$ and transverse fields $h_i$ are independent random variables
taken from distributions restricted to positive values, 
while $\epsilon_{h,i}$ and $\epsilon_{J,i}$ 
(also restricted to be positive)
parametrize the strength of the coupling between the colors.\footnote{Even if we assume uniform nonrandom values of $\epsilon_J$ and $\epsilon_h$,
they will acquire randomness under renormalization.}
Various versions of the AT model have been used to describe the layers
of atoms absorbed on surfaces, organic magnets, current loops in
high-$T_c$ superconductors as well as the elastic response of DNA
molecules.
Note the invariance of the Hamiltonian under the following duality transformation:
$\sigma_{\alpha,i}^z \sigma_{\alpha,i+1}^z \to \tau_{\alpha,i}^x$, 
$\sigma_{\alpha,i}^x \to \tau_{\alpha,i}^z \tau_{\alpha,i+1}^z$,
 $J_i\rightleftarrows h_i$, and $\epsilon_{J,i}\rightleftarrows\epsilon_{h,i}$, 
where $\tau^x$ and $\tau^z$ are the dual Pauli operators.
The bulk phases of the AT model (\ref{H_label}) are easily understood.
If the typical interaction $J_{typ}$ is larger than the typical field $h_{typ}$, 
the system is in the ordered (Baxter) phase in which each color orders ferromagnetically.
When $h_{typ}\gg J_{typ}$, the model is in the paramagnetic phase.
If there is a direct transition between these two phases, duality requires that
it occurs at $J_{typ}=h_{typ}$.
In the clean version of our system with $N\geq3$, the QPT between the paramagnetic
 and ordered (Baxter) phases is of 
first-order type.\cite{GrestWidom81,Fradkin84,Shankar85,Ceccatto91}

\section{Strong-Disorder Renormalization Group}
\label{SDRG}
To tackle the Hamiltonian~(\ref{H_label}), we now develop a SDRG method. 
In the weak-coupling regime ($\epsilon_h,\epsilon_J \ll \epsilon_c$, where 
$\epsilon_c$ is some $N$-dependent threshold), our method agrees with that of 
Goswami {\it et al.}\cite{GoswamiSchwabChakravarty08} Here, we focus on the strong
coupling regime $\epsilon_h,\epsilon_J \gg \epsilon_c$ where the method of the authors of
Ref.~\onlinecite{GoswamiSchwabChakravarty08} breaks down.

The basic idea of the SDRG method consists in identifying the largest local 
energy scale and perturbatively integrating out the corresponding 
high-energy degree of freedom. 
As we are in the strong-coupling regime,
this largest local energy is either a four-spin interaction (``AT interaction'') 
$k_{i}=\epsilon_{J,i}J_i$ or a two-color field-like term (``AT field'') 
$g_i=\epsilon_{h,i}h_i$. We thus define our high-energy
cutoff $\Omega = \max \{k_i,g_i\}$.


\begin{figure}
\includegraphics[width=0.9\columnwidth]{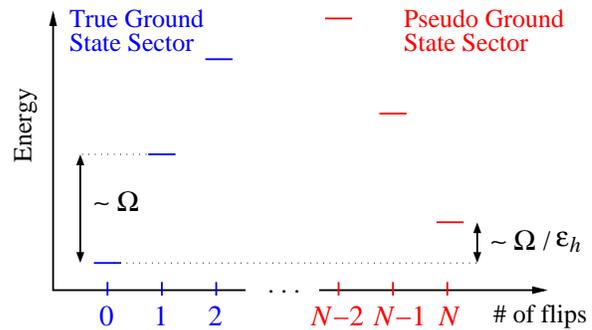}
\caption{(Color online) Spectrum of the unperturbed Hamiltonian (\ref{H0}) as function
of the number of colors flipped with respect to the ground state  $\left |\to , \to , \cdots , \to \right \rangle$.
As long as $T\lesssim\Omega/\epsilon_h$, the pseudo ground state $|\phi_0'\rangle=\left |\leftarrow , \leftarrow ,\cdots ,
\leftarrow \right \rangle$ can be neglected when 
computing observables (stage 1 of the RG). When $T\gtrsim\Omega/\epsilon_h$, $|\phi_0\rangle$ and $|\phi_0'\rangle$
become effectively degenerate implying that both states need to be taken into account (stage 2 of the RG).}
\label{spectrum}
\end{figure} 

We now derive the decimation procedure.
If the largest local energy is an AT field located, say, at site $2$,
the unperturbed Hamiltonian for the decimation of this site reads
\begin{equation}
\label{H0}
 H_0=-\frac{g_2}{\epsilon_{h,2}}\sum_{\alpha=1}^N{\sigma_{\alpha,2}^x}
-g_2\sum_{\alpha<\beta}{\sigma_{\alpha,2}^x \sigma_{\beta,2}^x}.
\end{equation}
The ground state (GS) of $H_0$ is 
$\left |\phi_0\right \rangle=\left |\to , \to , \cdots , \to \right \rangle$, with energy 
$E_0=-N g_2/\epsilon_{h,2}-N(N-1)g_2/2$, where
each arrow represents a different color. 
Flipping $n$ colors leads to $\binom{N}{n}$ degenerate excited states with energy
$E_n=E_0+2ng_2/\epsilon_{h,2}+2n(N-n)g_2$.
In the strong-coupling regime, $\epsilon_h \gg 1$, the state 
$\left |\phi_0'\right \rangle= \left |\leftarrow , \leftarrow ,\cdots ,
\leftarrow \right \rangle$ plays a special role. 
Its energy $E_0'=Ng_2/\epsilon_{h,2}-N(N-1)g_2/2$ differs from that of the 
true ground state only by the subleading Ising term
$E_0'-E_0=2Ng_2/\epsilon_{h,2}$ (see Fig.~\ref{spectrum}). 
It can thus be considered a ``pseudo ground state'' which may be important for a correct
description of the low-energy physics.
The true and pseudo ground states each have their own sets of
 low-energy excitations which we call the ground-state and 
 pseudo-ground-state sectors of low energy states.

The couplings of site 2 to its neighbors,
\begin{equation}
\label{Pertu:of:H}
\begin{gathered}
V=
-\frac{k_{1}}{\epsilon_{J,1}}\sum_{\alpha=1}^N{\sigma_{\alpha,1}^z
\sigma_{\alpha,2}^z}
-k_{1}\sum_{\alpha<\beta}{\sigma_{\alpha,1}^z\sigma_{\alpha,2}^z
\sigma_{\beta,1}^z\sigma_{\beta,2}^z}\\
-\frac{k_{2}}{\epsilon_{J,2}}\sum_{\alpha=1}^N{\sigma_{\alpha,2}^z
\sigma_{\alpha,3}^z}
-k_{2}\sum_{\alpha<\beta}{\sigma_{\alpha,2}^z\sigma_{\alpha,3}^z
\sigma_{\beta,2}^z\sigma_{\beta,3}^z},
 \end{gathered}
\end{equation}
is the perturbation part of the Hamiltonian.
We now decimate site 2 in the second-order perturbation theory, keeping both the true ground state 
and the pseudo ground state. It is important to note that second-order perturbation theory does not mix states
from the two sectors as long as $N > 4$. (The sectors are coupled in a higher order
 of perturbation theory, but these terms are irrelevant at our IRCP).
 After decimating site 2, the effective interaction Hamiltonian of the neighboring sites reads (in the large-$\epsilon_J$ limit)
\begin{equation}
\label{H1eff}
\tilde{H}_{eff} = 
-\frac{\tilde{k}}{\tilde{\epsilon}_J}\sum_{\alpha=1}^N{\sigma_{\alpha,1}^z
\sigma_{\alpha,3}^z}
-\tilde{k}\sum_{\alpha<\beta}{\sigma_{\alpha,1}^z\sigma_{\alpha,3}^z
\sigma_{\beta,1}^z\sigma_{\beta,3}^z}-\tilde{\omega}\tilde{\zeta},
\end{equation}
with 
\begin{equation}
\label{keff}
\tilde k=\frac{k_1k_2}{2(N-2)g_2},~
~~\tilde \epsilon_J=\frac{\epsilon_{J,1}\epsilon_{J,2}}{2}\frac{N-1}{N-2},~
~~\tilde{\omega}=Ng_2/\epsilon_{h,2}.
\end{equation}

\begin{figure}
\includegraphics[width=0.9\columnwidth]{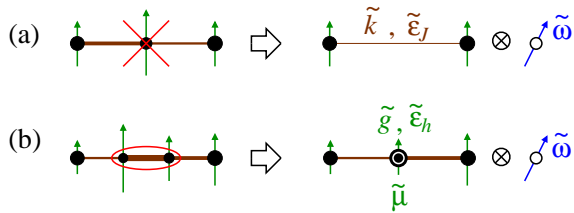}
\caption{(Color online) a) Decimating a site results in a renormalized bond (characterized by $\tilde k$ and $\tilde \epsilon_J$)
between its neighbors, and it introduces an extra binary ``sector'' degree of freedom represented as an Ising spin
$\tilde\zeta=\pm 1$ in an external field $\tilde \omega$ [see Eq.~(\ref{Heff})]. b) Decimating a bond results in a renormalized 
site characterized by $\tilde g$, $\tilde \epsilon_h$ and  another sector degree of freedom.
}
\label{decimation}
\end{figure}

Here, $\tilde{\zeta}=\pm 1$ is a new
Ising degree of freedom which represents the energy splitting between
the true and the pseudo ground states. In the large-$\epsilon_J$ regime, it is only very weakly coupled 
to the rest of the chain and can be considered free.
In Fig.~\hyperref[decimation]{\ref{decimation}(a)}, we sketch this decimation
procedure.

The decimation of a bond can be treated in the same way. If an AT four-spin
interaction, say $k_2$, is the largest local energy, the unperturbed Hamiltonian reads
\begin{equation}
H_0=-\frac{k_2}{\epsilon_{J}}\sum_{\alpha=1}^N{\sigma_{\alpha,2}^z\sigma_{\alpha,3}^z}
-k_2\sum_{\alpha<\beta}^N{\sigma_{\alpha,2}^z\sigma_{\alpha,3}^z\sigma_{\beta,2}^z\sigma_{\beta,3}^z}.
\end{equation}
Its GS is obtained by any sequence of parallel nearest-neighbors pairs
 (e.g. $|\phi_0\rangle=|\uparrow\uparrow,\uparrow\uparrow,\downarrow\downarrow,\downarrow\downarrow
,\uparrow\uparrow,\downarrow\downarrow,\cdots,\uparrow\uparrow\rangle$) with energy
 $E_0=-Nk_2/\epsilon_{J,2}-N(N-1)k_2/2$.
As above, in the strong-coupling limit $\epsilon_{J,2}\gg1$, $H_0$ has a pseudo-GS consisting of a sequence of anti-parallel nearest-neighbors pairs 
 (e.g. $|\phi_0'\rangle=|\uparrow\downarrow,\uparrow\downarrow,\downarrow\uparrow,
\downarrow\uparrow,\uparrow\downarrow,\downarrow\uparrow,\cdots,\uparrow\downarrow\rangle$)
 with energy $E_0'=Nk_2/\epsilon_J-N(N-1)k_2/2$.

When integrating out the bond, the two-site cluster gets replaced by a single site which contains one additional internal binary
degree of freedom, namely, whether the cluster is in the GS sector or in the pseudo-GS sector.
Its effective Hamiltonian reads
\begin{equation}\label{Heff}
\tilde H_{eff}=-\frac{\tilde g}{\tilde\epsilon_h}\sum_{\alpha=1}^{N}{\sigma_{\alpha,2}^x}-\tilde g\sum_{\alpha<\beta}{\tilde\sigma_{\alpha,2}^x\tilde\sigma_{\beta,2}^x}-\tilde\omega\tilde \zeta
\end{equation}
with
\begin{equation}\label{geff}
 \tilde g=\frac{g_2g_{3}}{2k_2[N-2]},~~~
 \tilde \epsilon_h=\frac{\epsilon_{h,2}\epsilon_{h,3}}{2}\frac{N-1}{N-2},~~~
 \tilde \omega=Nk_2/\epsilon_{J,2}.
\end{equation}
Here, $\tilde\zeta$ distinguishes the two sectors as before. The 
duality of the Hamiltonian can be seen by comparing Eqs. (\ref{keff}) and (\ref{geff}) after exchanging
the roles of $k$ and $g$ as well as $\epsilon_h$ and $\epsilon_J$.

Note that the magnetic moment $\tilde\mu$ of the new effective site 
depends on the internal degree of freedom $\tilde\zeta$ [see Fig.~\hyperref[decimation]{\ref{decimation}(b)}] 
because neighboring spins are parallel in the GS sector
but antiparallel in the pseudo-GS sector.
We will come back to this point when discussing observables.

The SDRG proceeds by iterating these decimations. In this process,
the coupling strengths $\epsilon_J$, $\epsilon_h$ flow to infinity if their initial
values are greater than some $\epsilon_c(N)$. 
This means that the Ising terms $J_i, h_i$ become less and less important with
decreasing energy scale $\Omega$. The large-$\epsilon$ approximation 
thus becomes asymptotically exact.
The remaining energies are the AT four-spin interactions $k_i$ and the AT fields $g_i$.
Their recursions relations have the same multiplicative structure as the recursions of
Fisher's solution \cite{Fisher95} of the random transverse-field Ising model.
The flow of the distributions $P(k_i)$, $R(g_i)$ and their fixed points are thus 
identical to those of Fisher's solution, see Fig.~\ref{FlowDiagram}.
We conclude that the distributions of $k$, $g$ have an infinite-randomness critical
fixed point featuring exponential instead of power-law scaling. 
\cite{Fisher92,Fisher95,RiegerYoung96} As the Ising coupling $J_i, h_i$ have vanished,
this critical fixed point has the symmetry of the AT interaction and field
 terms which is higher than that of the full Hamiltonian.

\section{Phase Diagram and Observables}
\label{PDO}
The zero-temperature phase diagram of our system is determined by the low-energy limit 
of the SDRG flow. 
There are three classes of fixed points parameterized by the distance from
criticality $r=\ln(g_{typ}/k_{typ})=\langle \ln g\rangle-\langle \ln k\rangle$
(where $\langle\cdots\rangle$ denotes the disorder average): 
The critical fixed point at $r=0$, and two lines of fixed points for 
the ordered ($r<0$) and for the disordered ($r>0$) Griffiths phases.
This implies that there is a direct continuous phase transition between the ordered 
(Baxter) and disordered phases.
We found no evidence for additional phases or phase transitions. In agreement with the 
Aizenman-Wehr theorem,\cite{GreenblattAizenmanLebowitz09} we thus 
conclude that disorder turns the clean first-order QPT into a continuous QPT in both
strong-coupling and weak-coupling regimes.
\begin{figure}
\includegraphics[width=0.9\columnwidth,height=0.55\columnwidth]{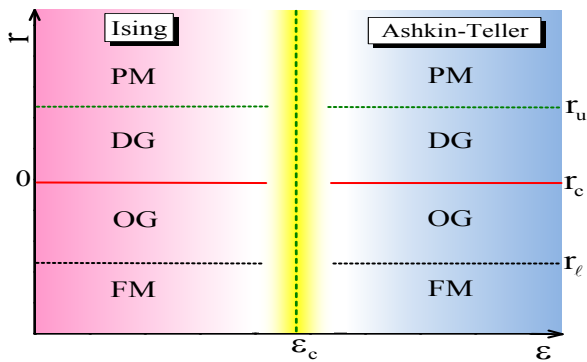}
\caption{(Color online) Phase diagram of the $N$-color quantum Ashkin-Teller model as function of $r=\ln(h_{typ}/J_{typ})$
 and the intercolor coupling $\epsilon$ at fixed disorder strength.
 The critical line is located at $r=r_c=0$ ($J_{typ}=h_{typ}$) as expected from the duality transformation. PM and FM denote the conventional 
paramagnetic and ferromagnetic (Baxter) phases. OG and DG denote the ordered and disordered Griffiths phases.}
\label{PDFig}
\end{figure} 
We now turn to the behavior of observables at low temperatures.
Let us fix the intercolor coupling parameter at some $\epsilon>\epsilon_c$ and tune the transition by
 the ratio $h_{typ}/J_{typ}=g_{typ}/k_{typ}$~(see Fig. \ref{PDFig}).
The basic idea is as follows.\cite{BhattLee82} 
We decimate the system until the cutoff
energy scale $\Omega$ reaches the temperature $T$. For low enough $T$, the
distributions of all energy scales in the renormalized system become very broad, 
and thus, the remaining degrees of freedom can be considered as free.
Applying this procedure, we have to distinguish two stages depending on the importance of the pseudo ground state.
(1) Both AT and Ising couplings are above the temperature.
(In this stage, we decimate sites and bonds whose internal sector degrees 
of freedom are frozen in the true ground state, $\tilde\zeta=1$.)
(2) The temperature is below the AT couplings but above the Ising couplings.
(Here we still decimate sites and bonds, but their internal degrees of freedom are 
free, i.e., they can be in either of the two sectors, $\tilde\zeta=\pm 1$.)

Let us illuminate this RG scheme on the example of the entropy.
\footnote{We will focus at low enough temperatures such that the RG flow
reaches the nontrivial second stage 
A detailed discussion for the high-temperature behavior including crossovers
will be given elsewhere\cite{HrahshehHoyosVojta_unpublished}}
When the RG flow stops at $\Omega=T$, all spins are completely free. A surviving cluster has $2^N$
available states (two per independent color) giving an entropy contribution of $N\ln2$,
i.e., $S_{\rm chain} = n_TN\ln2$, where
$n_\Omega$ is the density of surviving clusters at energy scale 
$\Omega$ (Ref.~\cite{Fisher95}).
Moreover, during stage 2 of the flow, residual entropy was accumulated in the
internal degrees of freedom, each of them contributing $\ln2$ to the entropy.
Noticing that each stage-2 RG decimation generates one extra
degree of freedom, and that stage 2 starts when $\Omega/\epsilon_{J,h} = T$, 
($\epsilon_x$ is the {\it typical} value of $\epsilon_{x,i}$
at energy scale $\Omega$),
the extra contribution to the entropy is 
$S_{\rm extra} = [w_J (n_{\epsilon_J T}-n_{T})+w_h (n_{\epsilon_h T}-n_{T})]\ln2$, 
with $w_{J}=1-w_{h}$ being the fraction of coupling decimations in the
entire stage 2 of the RG flow.
To compute $S_{\rm extra}$ we need to know how $\epsilon_J$ and $\epsilon_h$ depend on $\Omega$.
From the recursions (\ref{geff}) and (\ref{keff}),
 it is clear that $\ln \epsilon_h$ (and $\ln \epsilon_J$) scale like the number of sites (bonds) in a renormalized cluster (larger bond). 

At criticality, $w_J=w_h=1/2$, $n_\Omega \sim [\ln (\Omega_I/\Omega)]^{-1/\psi}$,
with $\psi=1/2$ being the tunneling exponent, and 
$\ln \epsilon_h = \ln \epsilon_J \sim [\ln(\Omega_I/\Omega)]^\phi$, with
$\phi = \frac 12 (1+\sqrt{5})$ (Refs.~\cite{Fisher95} and~\cite{VojtaKotabageHoyos09}). Thus, summing the two contributions we find that
\begin{equation}
 S=C_1\left[\ln{\left(\frac{\Omega_I}{T}\right)}\right]^{-\frac{1}{\psi\phi}}\ln2
+C_2\left[\ln{\left(\frac{\Omega_I}{T}\right)}\right]^{-\frac{1}{\psi}}N\ln2,
\end{equation}
where $C_1$ and $C_2$ are nonuniversal constants, and
$\Omega_I$ is the bare energy cutoff.
As $\phi>1$, the low-$T$ entropy becomes dominated by the extra degrees of
freedom $S\to S_{\rm extra} \sim [\ln(\Omega_I/T)]^{-1/(\phi\psi)}$.

In the ordered Griffiths phase ($r<0$), $w_J\to1$ and
$\ln\epsilon_J=Az^{\nu\psi(\phi-1)}\ln(\Omega_I/\Omega)$,
with $A$ being a nonuniversal constant of order unity, 
$\nu = 2$ the correlation length exponent, and
$z=1/(2|r|)$ the dynamical exponent.
As $n_\Omega \sim |r|^{\nu}(\Omega/\Omega_I)^{1/z}$, we find that
\begin{equation}
\label{Sextraordered}
S_{\rm extra} \sim |r|^\nu (T/\Omega_I)^{1/(z+Az^\phi)}\ln2,
\end{equation}
 which dominates over
the chain contribution proportional to $ T^{1/z}N\ln2$. As expected from duality,
the same result holds for the disordered phase ($r>0$).

To discuss the magnetic susceptibility, we need to find the effective magnetic 
moment $\mu_{\rm eff}$ of a cluster surviving at the RG energy scale $\Omega=T$.
If all internal degrees of freedom were in their ground state, 
$\mu_{\rm eff}$ would be given by the number of sites in the cluster. 
However, analogously to the entropy, $\mu_{\rm eff}$ is modified because
of the stage 2 of the RG flow. 
In this stage, the internal degrees of freedom are free, meaning not all spins 
in a surviving cluster are parallel, reducing the effective moment. 
A detailed analysis based on the central limit theorem 
\cite{HrahshehHoyosVojta_unpublished} gives
$\mu_{\rm eff}\sim [\ln(\Omega_I/T)]^{\phi/2+1/2}$ at criticality and
$\mu_{\rm eff}\sim r^{\nu\psi(1-\phi)}[\ln(\Omega_I/T)]^{1/2}$ 
in the disordered Griffiths phase, as well as 
$\mu_{\rm eff}\sim r^{-\phi/2}T^{-1/(2z)}$ in the ordered Griffiths phase.

The magnetic susceptibility $\chi(T)$ can now be computed.
All eliminated clusters had AT fields greater than the temperature, 
and thus do not contribute to $\chi$ since they are fully polarized in the $x$-direction,
whereas the surviving clusters are effectively free and contribute with a 
Curie term: $\chi \sim \mu_{\rm eff}^2 n_T/T$.
We find that
\begin{equation}
\label{chiTc}
\chi\sim [\ln(\Omega_I/T)]^{\phi+1-1/\psi}/T
\end{equation}
 in the critical region,
while it becomes 
\begin{equation}
\label{chiDG}
\chi\sim r^{\nu+2\nu\psi(1-\phi)}T^{1/z-1}\ln(\Omega_I/T)
\end{equation}
in the disordered Griffiths phase, and 
take a Curie form $\chi\sim |r|^{\nu-\phi}T^{-1}$ in the ordered Griffiths phase.

\section{Conclusion}
\label{Conc}
In summary, we have solved the random quantum Ashkin-Teller model by means of a
strong-disorder renormalization-group method that works not just for weak-coupling but
also in the strong-coupling regime and yields asymptotically exact results. 
In the concluding paragraphs, we put our results into broader perspective. 

First, we have demonstrated that random disorder turns the first-order QPT 
between the paramagnetic and Baxter phases into a continuous 
one not just in the weak-coupling regime but also in the strong-coupling regime. 
This resolves the seeming contradiction between
the quantum Aizenman-Wehr theorem 
\cite{GreenblattAizenmanLebowitz09, GreenblattAizenmanLebowitz12} 
and the conclusion that the first-order transition
may persist for sufficiently large coupling strength.\cite{GoswamiSchwabChakravarty08}

The resulting continuous transition is controlled by two different IRCPs in 
the weak and strong coupling regimes. For weak coupling, the critical
point is in the universality class of the random transverse-field Ising chain.
\cite{Fisher95} 
For strong coupling, we find a distinct type of IRCP 
which features a higher symmetry than the underlying Hamiltonian.
The associated internal degrees of freedom lead to
 even stronger thermodynamic singularities both at criticality and in the 
Griffiths phases.

Our results apply to $N>4$ colors where the true and pseudo ground-state sectors 
are not coupled.
As a result, the Ising terms in the Hamiltonian are irrelevant perturbations 
(in the renormalization group sense) at our IRCP.
The case $N\leq4$ is special because the two sectors get coupled and thus requires a 
separate investigation.
Interestingly, novel behavior has been recently verified for the 
{\it classical} transition in the two-dimensional AT model \cite{BKTC12} for $N=3$.

Our explicit calculations were for one space dimension. However, we believe that
many aspects of our results carry over to higher dimensions. In particular, the
SDRG recursion relations take the same form in all dimensions (as they are purely local).
This implies that the RG flow for large inter-color coupling $\epsilon$ will
be toward $\epsilon\to \infty$ as in one dimension. Moreover, the flows of the AT
energies $g$ and $k$ (although not exactly solvable in $d>1$) are identical
to the flows of the random transverse-field Ising model in the same dimension.
In two and three dimensions, these flows have been studied numerically,\cite{MMHF00,IgloiMonthus05,KovacsIgloi11}
yielding IRCPs as in one dimension.
We thus conclude that the strong-coupling regime of the random quantum AT model will be 
controlled by an Ashkin-Teller IRCP not just in one dimension but also in two and three dimensions.  

We note that our method is also interesting from a general 
renormalization-group point of view.
After a decimation, the resulting system cannot be represented
solely in terms of a renormalized quantum AT Hamiltonian because 
the internal degree of freedom needs to be taken into account.
Normally, the appearance of new variables dooms an RG scheme.
\footnote{Or it requires a generalization that includes all new terms in 
the starting Hamiltonian}
Here, however, the new variables, despite their influence on observables, 
are inert in the sense that they do not influence the RG flow of the other 
terms in the Hamiltonian, which makes the problem tractable.
We expect that this insight may be applicable to renormalization-group schemes 
in other fields.

The strong-disorder RG approach to the random quantum AT model gives asymptotically exact results for both sufficiently weak and sufficiently strong
coupling ($\epsilon\ll 1$, $\epsilon\gg 1$), see Fig.~\ref{FlowDiagram}. The behavior for moderate $\epsilon$ is not exactly solved. In the simplest scenario, the weak-coupling
and strong-coupling IRCPs are separated by a unique multicritical point at some
 intermediate coupling, however, more complicated scenarios cannot be excluded.
The resolution of this question will likely come from numerical implementations of the SDRG and /or (quantum) Monte Carlo simulations. 

This work has been supported in part by the NSF under Grants No. DMR-0906566 and DMR-1205803, 
by FAPESP under Grant No. 2010/ 03749-4, and by CNPq
under Grants No. 590093/2011-8 and No. 302301/2009-7.
                                                          
\bibliographystyle{apsrev4-1}
\bibliography{rareregions}

\end{document}